# Dynamic heterogeneity in the self-induced spin glass state of elemental neodymium

**Authors:** L. Niggli[1], J. H. Strik[1], Z. Liu[1], A. Bergman[2], M. I. Katsnelson[1], D. Wegner[1], A. A. Khajetoorians[1]*

**Affiliations:**

[1]Institute for Molecules and Materials, Radboud University, Nijmegen, The Netherlands

[2]Department of Physics and Astronomy, Uppsala University, Uppsala, Sweden

*corresponding author: a.khajetoorians@science.ru.nl

**Abstract:** Spin glasses are magnetic materials exhibiting numerous magnetization patterns, that randomly vary both in real space and in time. To date, it is still not well understood what the nature of these spatiotemporal dynamics is, namely if they are completely random or if there are links between given time and length scales. Here we show the ubiquitous behavior of dynamic heterogeneity in the self-induced spin glass state of elemental neodymium. We used spin-polarized scanning tunneling microscopy in combination with atomistic spin dynamics simulations to image the locally ordered magnetic patterns in the glass state, and tracked the induced spatiotemporal dynamics in response to external perturbations. We observed that the real space magnetization exhibited a coexistence of slow and fast dynamics reminiscent of dynamic heterogeneity in structural glasses. Furthermore, we found that zero-field cooling imprints a specific set of metastable periodicities into the spin glass, which evolved during aging and could be thermally reinitialized. These results demonstrate the importance of local length scales for the understanding of aging dynamics in spin glasses and provide a link to the more general picture of true glasses.

**Main Text:** The complexity of spin glasses continues to be the subject of intense experimental and theoretical interest. Its magnetic state is often described as a distribution of seemingly random patterns that continuously evolve (*1, 2*). These dynamics, often referred to as aging, are characterized by the coexistence of multiple relaxation times spanning different orders of magnitude, where the magnetic state at a given time depends on its history, or age (*2, 3*). Aging dynamics distinguishes spin glasses from other frustrated magnets due to its inherent non-ergodic behavior (*1, 4*). It is usually linked to the randomization of exchange interactions, which is driven by disorder. The theoretical description of spin glasses is mostly based on various Ising spin models with competing long-range interactions (*5, 6*), such as the Sherrington-Kirkpatrick (SK) model which extends the coupling to infinite range (*7*). This mean-field-type problem was solved using the replica method (*4, 8-10*). The solution is length invariant, i.e. there are no favorable length scales that define particular domains composed of distinct patterns.

Prototypical spin glass materials show behavior where certain length scales and short-range order emerge (*11-14*), and these length scales may feature distinctive dynamics. It was shown that memory and rejuvenation can be retained during different aging cycles (*15, 16*), and has been described considering hierarchies of metastable states with different energetic favorability (*17-19*). These observations have led to a long-standing debate about the role of local length scales in spin glass materials and motivate descriptions beyond the mean-field limit. Experimentally, it is often challenging to disentangle the role of disorder on aging dynamics and the emergence of favorable length scales. To this end, it was shown that spin glass behavior can exist solely driven by frustration without disorder (*20-23*). This concept of a self-induced spin glass was demonstrated to explain the magnetism of elemental and crystalline neodymium (*24, 25*) and provides a platform to study spatiotemporal dynamics and the role of local length scales.

Here we demonstrate dynamic heterogeneity in the self-induced spin glass state of neodymium. Dynamic heterogeneity (DH) is a ubiquitous phenomenon seen in amorphous solids (*26*), where there is a correlation between distinct time scales and particular length scales. In amorphous solids, these spatiotemporal dynamics can lead to complex behavior, such as nucleation in real space of regions with correlated behavior (*27-31*). These descriptions go beyond the mean-field picture, and in spin systems have been predicted to affect the dynamics near the spin glass transition (*32-34*). It has been shown that this requires methods that incorporate real space and time information (*35, 36*). To date, observing DH in spin glasses has been extremely challenging: (i) due to the interplay between external disorder and the emergence of particular length scales, (ii) the time and length scales associated to DH in a spin glass, are often much faster and smaller, respectively, compared to amorphous solids.

**Self-induced spin glass in neodymium**

Using atomic-scale imaging to resolve aging dynamics in real space we explore DH in elemental neodymium. Its self-induced spin glass state is characterized by the coexistence of multiple and distinct locally ordered patterns in real space that can be imaged at the Nd(0001) surface with spin-polarized scanning tunneling microscopy (SP-STM), as seen in Fig. 1A (see methods and Ref. (*24*) for details). Each local pattern had no well-defined domain boundary, and could therefore not easily be delineated from other neighboring patterns. In Fourier space (Fig. 1B), the periodicities of these local patterns were derived from a spectrum of nearly degenerate $Q$-states, or so-called $Q$-pockets, that resulted from the frustrated exchange interactions, where $|Q| = 2\pi/\lambda$ denotes a magnetic wave vector (*24*). The fast Fourier transform (FFT) of the real-space magnetization *M*(*r*) is closely related to the static magnetic structure factor *S*(*Q*), as previously shown (*24*), and will be referred to like that for brevity. The *S*(*Q*) in Fig. 1B exhibited three distinct distributions of degenerate *Q*-states along each high-symmetry axis, which we refer to as $Q_A$-, $Q_B$- and $Q_C$-pockets, respectively (radially outwards). In the subsequent discussion, we study the evolution of magnetic patterns, labeled {$dQ_i$} to represent the whole set of patterns, in both real and Fourier space as a function of various thermodynamic variables, to identify as well as quantify DH.

**Snapshots of the dynamics**

To link the time-dependent behavior with local variations of the magnetization in real space, we imaged the frozen spatially-dependent magnetization after inducing dynamics through particular thermodynamic perturbations. We adopted the following 'dynamical procedure' as illustrated in Fig. 1C: (I) A given sample was zero-field cooled (ZFC) below its Néel temperature ($T_N$ = 19.9 K), through its long-range ordered multi-*Q* phase (*25*), and subsequently below the glass transition temperature ($T_g \sim 8$ K) down to *T* = 1.3 K. At *T* < $T_g$, long-range magnetic order dissolved into locally ordered magnetic patterns {$dQ_i$}, which exhibited complex aging dynamics as previously observed in Refs. (*24, 25*). These dynamics froze out at low temperatures and the various frozen {$dQ_i$} could be imaged at *T* = 1.3 K (and *B* = 0 T). At this temperature, no substantial changes in the magnetization over the course of 27 hours were observed (see Fig. S1). (II) We applied magnetic field cycles to *B* = 2-7 T to induce magnetization dynamics during the *B*-field sweep (*24*). The dynamics were frozen again when the applied field was removed, and the resulting evolved state could be re-imaged. Hence, by repeating step (II) we obtained a series of snapshots of the magnetization dynamics. (III) Lastly, after repeated *B*-field cycling and zero-field imaging, we warmed up the sample above $T_g$ to restart the procedure. We note that changes in temperature were not used to induce dynamics due to the presence of the multi-$Q$ state above $T_g$ (25). As previously shown, the trend with increasing temperature is to lower the entropy of the system.

**Evolution and reinitialization of $S(Q)$**

Fig. 2 illustrates an example of the dynamical procedure outlined in Fig. 1C to demonstrate the main findings. First, applying step (I) froze the self-induced spin glass into a particular state for a given film characterized by a distinct initial magnetic structure factor $S_0(Q)$ (Fig. 2A, D). Second, after repeating step (II) multiple times (10 magnetic field cycles '(II)$_{10}$' at $T$ = 1.3 K to $\Delta B$ = 7 T) an evolution of the $Q$-pockets was observed in the magnetization image in Fig. 2B resulting in a different final structure factor $S_{10}(Q)$ (Fig. 2E). It was characterized by both a different spatial distribution of magnetic patterns, i.e. the patterns have aged, but also a change of the periodicities themselves, hence $S_{10}(Q) \neq S_0(Q)$. Third, after performing step III and re-freezing the system (step (I)), we observed that thermal cycling of the system above $T_g$ reinitialized the self-induced spin glass (Fig. 2C, F), i.e. $S'_0(Q) \approx S_0(Q)$, whereas the specific spatial distribution of patterns $\{dQ_i\}$ was not reproduced (cooling rate dependence discussed in S3). The differences between $S_0(Q) \leftrightarrow S_{10}(Q)$ and $S_0(Q) \leftrightarrow S'_0(Q)$ are illustrated in Fig. 2G and 2H. Comparing $S_0(Q)$ and $S_{10}(Q)$, revealed a significant shift of spectral weight radially along the high-symmetry axes, corroborating a change of the periodicities induced by $B$-field cycling. In contrast to the comparison of $S(Q)$ during $B$-field cycling, there was significantly more overlap between the initial and the reinitialized states.

A detailed investigation of the various snapshots for a given sample, independent of the total cycle (I-III), showed that there was a systematic and continuous evolution of $S(Q)$ in response to the magnetic field cycles that was reproducible and not random. The evolution of the $Q$-pockets along the three high-symmetry axes for 10 field cycles to $\Delta B = 7$ T is illustrated as a sketch in Fig. 3A and the experimental data is shown in Fig. 3B. The $Q_A$-pocket shifted monotonically radially outwards redistributing its weight along the high-symmetry axis, while the $Q_B$- and $Q_C$-pockets shifted radially inwards. To quantify the saturation and reinitialization behavior of $S(Q)$, we calculated the Jensen-Shannon divergence $\mathcal{D}_{JS}$, see methods and Ref. (25), to obtain a measure of the similarity between a given distribution of $Q$-states, i.e. a particular $S(Q)$, at each point in the dynamical procedure. Based on our definition, maximally dissimilar distributions correspond to a $\mathcal{D}_{JS}$ value of 1, and identical distributions correspond to a value of 0. First, we evaluated the difference between consecutive magnetic structure factors during repeated field cycles by comparing $S_i(Q)$ to $S_{i+1}(Q)$ for different samples. The results are shown in Fig. 3C, where we observed a decrease of the value of $\mathcal{D}_{JS}$ during the $S(Q)$-evolution indicating that consecutive $S_i(Q)$ became more alike, and an apparent saturation of $\mathcal{D}_{JS}$ after 2-3 field cycles. Moreover, continued changes in local magnetic patterns $\{dQ_i\}$ in real space (see video S1) occurred even after the $S(Q)$-evolution saturated indicating that the material continues to age.

Second, we quantified the similarity of subsequent (re)initialized structure factors $S_0^i$ shown in Fig. 3D (green) as well as between (re)initialized $S_0^i$ and final states $S_{\geq 9}^i$ (blue) measured on the same sample. Together this resulted in a bimodal distribution in which there was a significantly larger dissimilarity between all (re)initialized and final states compared to all states resulting from ZFC below $T_g$ that have not yet seen a magnetic field after entering the spin glass phase. This implies that there was an inherent similarity between the sets of favored patterns $\{dQ_i\}$ associated with $S_0(Q)$ that were frozen in after ZFC cooling. Additionally, the systematic evolution of $S(Q)$ persisted for all the samples studied albeit with some variance in the given $S_0(Q)$ for each sample (see Fig. S4). The systematic changes together with the initialization indicated that the evolution of $S(Q)$ was not random, but followed a particular trend, independent of the sample. This suggests that the initial state, which was zero-field cooled, may be an intermediate metastable state.

**Real-space imaging of DH**

After ascertaining the aging dynamics and the evolution of $S(Q)$, we next illustrate dynamic heterogeneity by analyzing the changes in local magnetization patterns. Fig. 4A,B display two magnetization images obtained before ($t = t_0$) and after ($t = t_1$) applying a magnetic field cycle after $S(Q)$ has approximately saturated (third step (II) with a field value of $\Delta B = 2$ T). Randomly sampling both images revealed spatial regions in which there were no changes to a given pattern (examples in Fig. 4H) as well as spatial regions where the pattern had changed (see Fig. 4G). The coexistence of both similar and dissimilar regions indicated the presence of heterogenous dynamics, namely *slow* (no change) and *fast* (change) dynamics linked to particular regions in real space.

To investigate the spatial distribution of the identified slow and fast dynamics more systematically, we divided both magnetization images into an array of $12 \times 12$ nm$^2$ boxes, each containing approximately 1000 spins. In Fig. 4C, we color-coded the individual boxes in the original image (Fig. 4A) to illustrate three types of boxes: (a) slow dynamics/stable pattern (blue), (b) fast dynamics/changing pattern (green) and (c) undetermined (transparent). A detailed discussion of this qualitative sorting analysis can be found in the supplementary text, section S5. There was a distribution of regions linked to both slow and fast time scales. This behavior persisted for continued cycling also with higher magnetic fields, as evident in Fig. 4D-F and Fig. S6. In all these examples, we did not observe the dominance of one or more sets of $Q$-states, as would be expected for the growth of a favorable domain. In addition, we observed no correlation with defects (located at the bright spots in the magnetization images), which could be imaged in the absence of spin contrast at other voltages. We note that this behavior could be observed for variable box sizes $L$ (Fig. S7) as long as $L$ remained smaller than the average size of the patterns. We also note that spatiotemporal dynamics have been extensively studied in amorphous solids using the concept of a real-space four-point correlation function (*35, 36*). This function, together with its associated generalized susceptibility, is used to quantify the typical size of dynamically correlated regions (*37-39*). However, it is challenging to apply this concept directly to the magnetic patterns in neodymium as it is unclear how to link the multi-$Q$ nature of the local spin-spin correlation to the conventional definition of a four-point correlation function.

In order to further substantiate the experimental observation of DH, we performed atomistic spin dynamics (ASD) simulations in which we compared subsequent snapshots of the spatially dependent magnetization. Fig. 5A shows such a simulated real-space magnetization image at an initial time $t_0 = 1$ ps (see Fig. S8 for a decomposition into the two magnetic sublattices and S9 for more details). The magnetization revealed multiple types of local magnetic order of multi-$Q$ character, without any favorable long-range multi-$Q$ state. We subsequently tracked the evolution of the magnetization over time and show in Fig. 5B one snapshot during the evolution, after $t = 500$ ps (see also video S2). Clearly, particular regions in the image had different patterns $\{dQ_i\}$, compared to Fig. 5A. Simultaneously, there were areas in the image which have not changed. To illustrate these changes, we used a subtraction method between the two images (supplementary text, section S10), to highlight the regions that have changed compared to those that have not (Fig. 5C). This provided evidence for dynamic heterogeneity. However, based on these two images alone, it cannot be substantiated if these fluctuations were due to random switching, or if there was a systematic evolution of the real space magnetization patterns that were linked to the various time scales.

We subsequently illustrated the evolution of the magnetization at various time steps ($\Delta t = 70$ ps) to show nucleation behavior associated with DH. In Fig. 5D, we identified and color-coded the regions of change during the evolution (subtraction method in S9), comparing each subsequent time step with the original magnetization pattern. The image showed that the evolution of the changes in the magnetization patterns was neither consistent with stochastic behavior (i.e. random switching) nor with the clear movement of a favorable domain. It rather showed the emergence of regions with particular dynamic behavior. We observed analogous behavior for all investigated samples (see Fig. S11) indicating that the development of dynamically correlated regions was inherent to elemental neodymium. This is reminiscent of DH seen in structural glasses, where particles with similar mobilities cluster together in space and the relaxation times among these different regions can vary by orders of magnitude (*30, 40*). Accordingly, the self-induced spin glass state of neodymium possesses particular length scales inherent to both the real-space structure as well as the dynamics of the magnetization, unlike the expected random uniformity linked to the mean-field description of spin glasses. We note that while the numerical simulations reproduced the observed spatially heterogeneous dynamics and demonstrated glassy dynamics in the two-time autocorrelation function (see Fig. S12), they did not find evidence for a change of the magnetic structure factor.

**Conclusion**

We have studied, both experimentally and theoretically, the spatially resolved magnetization dynamics of the self-induced spin glass state in neodymium and demonstrated dynamic heterogeneity. This observation of DH demonstrates that this ubiquitous phenomenon seen in amorphous solids also exists in a spin glass system, necessitating descriptions beyond the mean-field limit. It creates a natural bridge between the two different paradigms, and future studies can also harness approaches to quantify the role of local length scales (*35, 36*). Moreover, it suggests that local regions may age toward equilibrium at different rates as theoretically explored in studies on non-local aging (*33, 41*) and that local length scales have to be taken into account to construct a mathematically rigorous theory of nonequilibrium spin glass dynamics (*42*). It remains to be seen how the spatiotemporal dynamics and the specific length scales introduced in the problem are linked to the favorability of the various periodicities ($Q$-states) in the material. In addition, we observed an evolution of the magnetic structure factor of neodymium in the self-induced spin glass state during aging dynamics induced by magnetic field cycles. This suggests metastable behavior upon zero-field cooling. The origin of this effect is not clear, but it may be linked to the high-temperature multi-$Q$ state present in neodymium. This provides a pathway to study analogous concepts to phase-change memory near phase transitions.

**Acknowledgments:** We thank O. Eriksson and A. Mauri for fruitful discussions.

**Funding:** This project was supported by the European Research Council (ERC) under the European Union's Horizon 2020 research and innovation program (grant no. 818399). This publication is part of the project "What can we 'learn' with atoms?" (with project number VI.C.212.007) of the research program VICI which is (partly) financed by the Dutch Research Council (NWO). This publication is part of the 'Self-induced spin glasses – a new state of matter' project (OCENW.KLEIN.493) of the KLEIN research program which is (partly) financed by the Dutch Research Council (NWO). A.A.K. and J.H.S. acknowledge the research program "Materials for the Quantum Age" (QuMat) for financial support. This program (registration number 024.005.006) is part of the Gravitation program financed by the Dutch Ministry of Education, Culture and Science (OCW). M.I.K acknowledges the European Research Council via Synergy grant number 854843 (FASTCORR). A.B. acknowledges support from eSSENCE and the Carl Tryggers foundation. The computations were enabled by resources provided by the National Academic Infrastructure for Supercomputing in Sweden (NAISS), partially funded by the Swedish Research Council through grant agreement no. 2022-06725.

**Author contributions:** L.N., J.H.S, and Z.L. performed the experiments. L.N. and J.H.S. performed the experimental analysis. L.N., J.H.S., A.B., M.I.K., D.W., and A.A.K. designed the analysis methods and all authors participated in the iterative discussions about the analysis and its scientific conclusions. A.B. performed the ASD simulations. L.N., M.I.K., D.W., and A.A.K. designed the experiments. L.N. and A.A.K. primarily wrote the manuscript, while all authors provided input during its development.

**Competing interests:** The authors declare no competing interests.

**Data and materials availability:** All data needed to evaluate the conclusions in the paper are present in the paper or the Supplementary Materials.

**Supplementary Materials:**

Materials and Methods

Supplementary Text

Figs. S1 to S12

References (*43-46*)

Movies S1 to S2

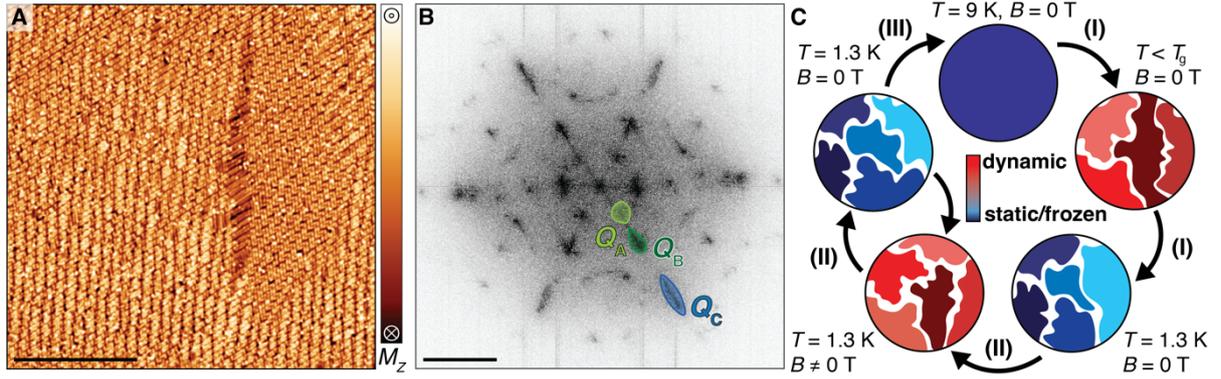

**Fig. 1. The self-induced spin glass state of Nd(0001) and the dynamical cycle.** (A) Frozen real space magnetization image $M(r)$ measured at $B = 0$ T, $T = 1.3$ K (zoom of Fig. 2A) illustrating local order and (B) $S(Q)$ obtained by fast Fourier transformation of Fig. 2A. $Q$-pockets along one high-symmetry axis, labeled $Q_A$, $Q_B$, $Q_C$, are highlighted (image parameters: $I_t = 200$ pA; scale bar $M(r)$: 50 nm, scale bar $S(Q)$: 3 nm$^{-1}$, color scale for $S(Q)$-images is a linear gradient). (C) Diagram illustrating the dynamical procedure as a function of temperature $T$ and magnetic field $B$: (I) Starting from long-range magnetic order ($T > T_g \approx$ 8 K; multi-$Q$ domain in dark blue (static)), zero-field cooling below $T_g$ leads to emergence of local order exhibiting aging dynamics (shades of red) that are effectively frozen out at $T = 1.3$ K (shades of blue). (II) Magnetization dynamics are induced during out-of-plane magnetic field cycles and frozen again when the field is reduced to zero. (III) The system is reinitialized through temperature cycling via the ordered state (and (I)) back to $T = 1.3$ K.

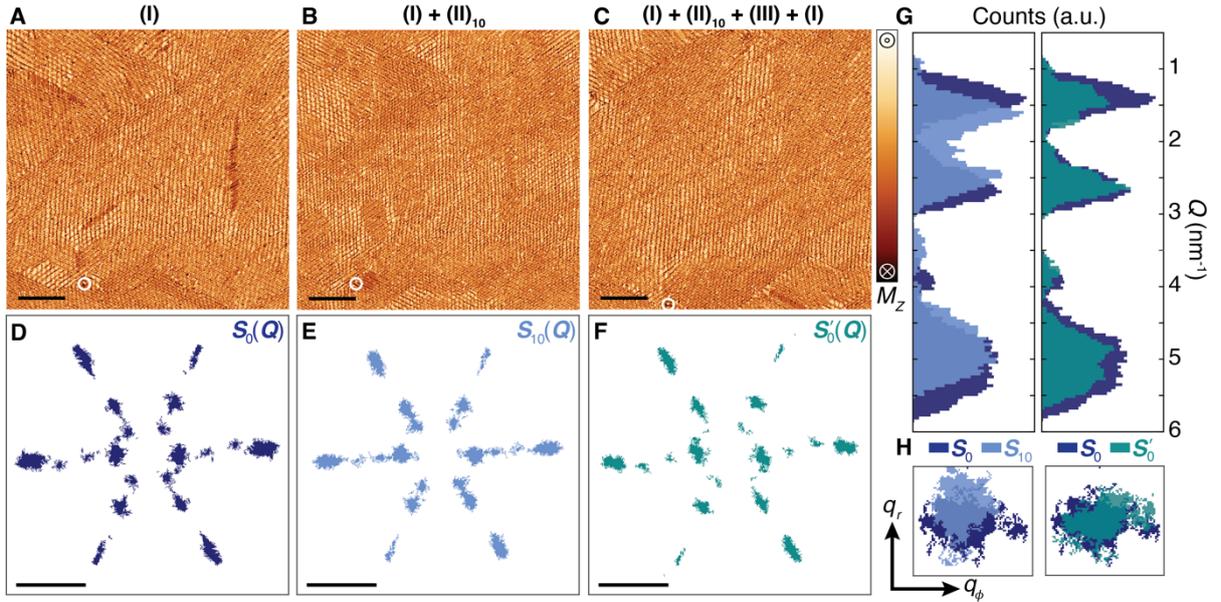

**Fig. 2. Evolution of the magnetic structure factor during aging dynamics. (A-C)** Large scale magnetization images $M(r)$ measured at $B = 0$ T, $T = 1.3$ K and **(D-F)** corresponding filtered structure factors $S(Q)$ (details in S2). White circles in $M(r)$ mark the position of a defect as a common reference for all images. **(A, D)** Initial state $S_0$ reached after zero-field cooling (step (I)). **(B, E)** Final state $S_{10}$ reached after repeated $B$-field cycling (step (II) with $\Delta B_i = 7$ T, $i = 10$) with modified $Q$-pocket positions. **(C, F)** Reinitialization of structure factor (step (III)) into $S'_0$, which is similar to the initial structure factor $S_0 \approx S'_0$. **(G) (left)** Radial $Q$-distributions of initial $S_0$ (dark blue) and final $S_{10}$ (light blue) illustrate a change of the $Q$-pocket positions; **(right)** radial $Q$-distribution of $S'_0$ (teal) shows reinitialization of the $Q$-pocket positions into a $Q$-distribution matching $S_0$ (dark blue). **(H)** Zoom in of $S(Q)$ on the horizontal $Q_A$-pocket illustrating the difference between $S_0$ and $S_{10}$ as well as the similarity of $S_0$ and $S'_0$ (image parameters: $I_t = 200$ pA; scale bar $M(r)$: 50 nm, scale bar $S(Q)$: 3 nm$^{-1}$).

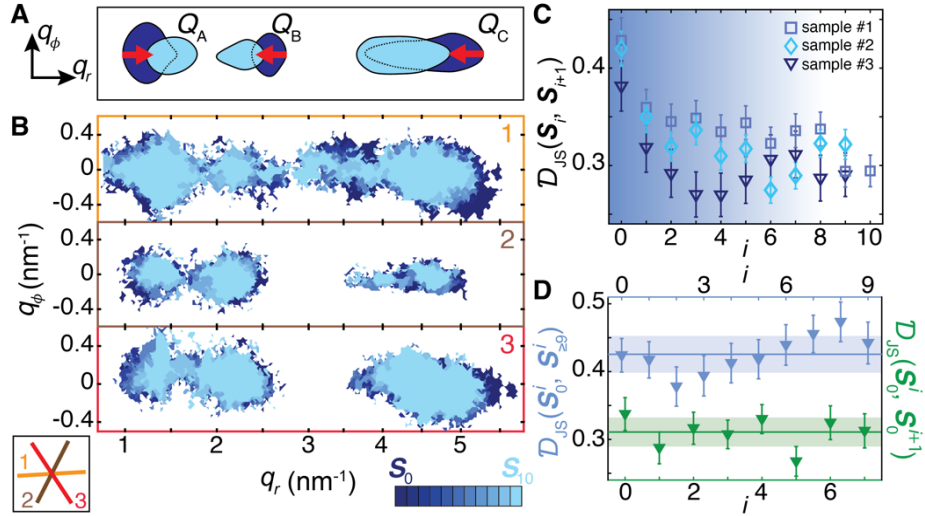

**Fig. 3. Quantification of the systematic evolution of $S(Q)$.** (**A**) Sketch of the $Q$-pockets along one high-symmetry direction before (dark blue) and after (light blue) repeated $B$-field cycles with red arrows indicating the radial positional change of each pocket. (**B**) $Q$-pockets along the three high-symmetry directions (labelled 1, 2, 3), extracted from $S_i(Q)$ during $i = 10$ field cycles to $B = 7$ T at $T = 1.3$ K, evolved in a structured way (dark to light blue); filtering method discussed in S2. (**C**) Quantification of the similarity between consecutive structure factors $S_i$ for different samples using the Jensen-Shannon divergence $\mathcal{D}_{JS}$ showing a reduction and saturation at a finite value of $\mathcal{D}_{JS}$. (**D**) Repeated evolution and reinitialization cycles for one particular sample (steps (II-III)) revealed a bimodal distribution in the divergence values, stemming from similarity ($\mathcal{D}_{JS} = 0.31 \pm 0.02$) of all reinitialized states (green) as well as a significant difference ($\mathcal{D}_{JS} = 0.43 \pm 0.03$) of all final and reinitialized states (light blue). The error bars are defined according to S2.

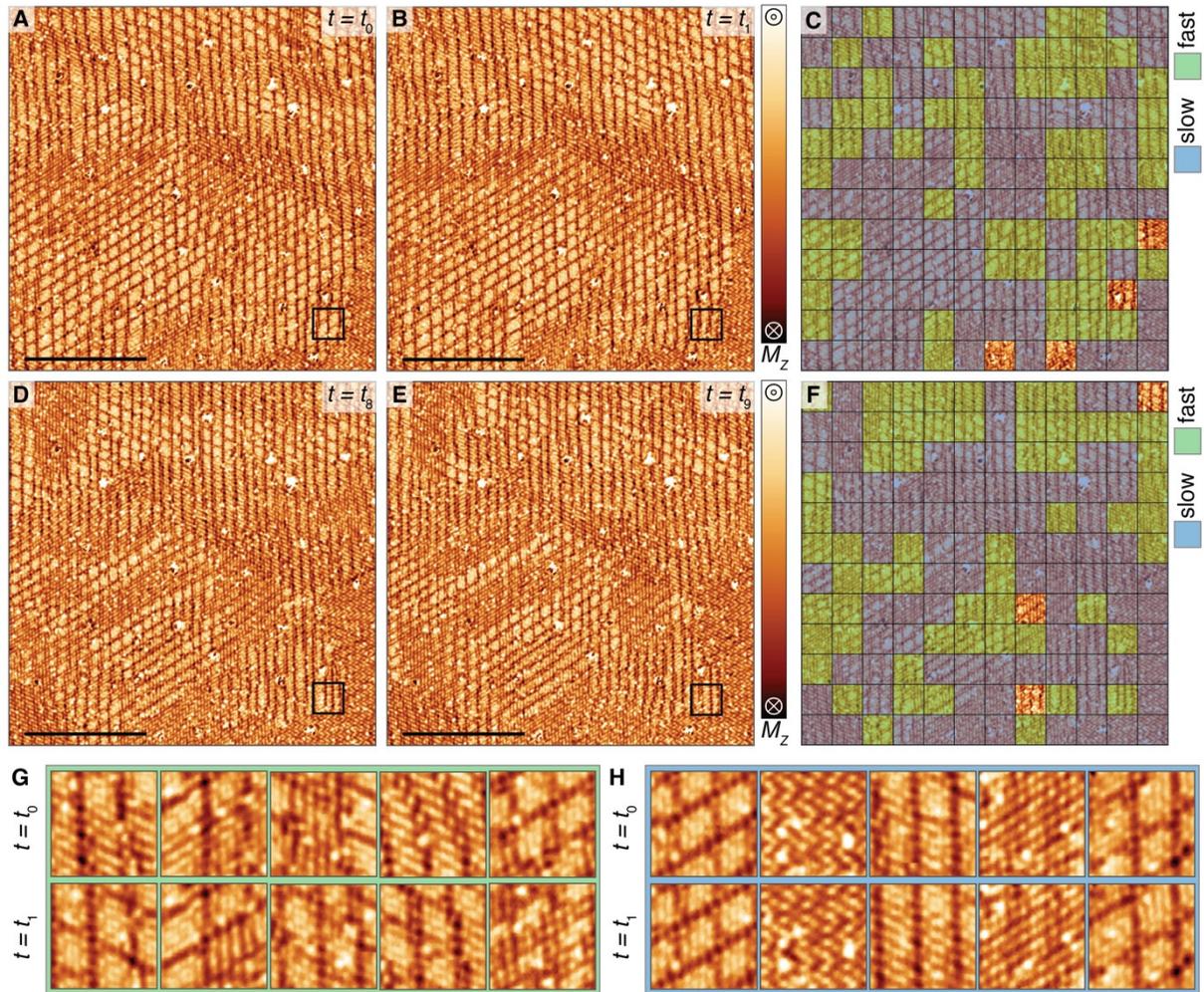

**Fig. 4. Local spatiotemporal dynamics and dynamic heterogeneity in magnetization images.** **(A-B)** Magnetization images before ($t = t_0$) and after ($t = t_1$) the third magnetic field cycle from B = 0T to $B$ = 2 T. **(C)** Spatial map of slow (blue) and fast (green) dynamics determined by eye from comparing patterns $dQ_i(t)$ for each box between $t_0$ and $t_1$ illustrating spatially heterogeneous dynamics. **(D-E)** Magnetization images before ($t = t_8$) and after ($t = t_9$) the 7$^{th}$ magnetic field cycle to $\Delta B$ = 6 T (11$^{th}$ cycle in total). **(F)** Qualitative spatial map of slow (blue) and fast (green) dynamics between $t_8$ and $t_9$. **(G-H)** Examples of different patterns possessing fast **(G)** and slow **(H)** dynamics (image parameters: $T$ = 1.3 K, $B$ = 0 T; $I_t$ = 200 pA; scale bar **(A-B, D-E)**: 50 nm; image sizes **(G-H)**: 12 nm, same colorbar as in **(A)**).

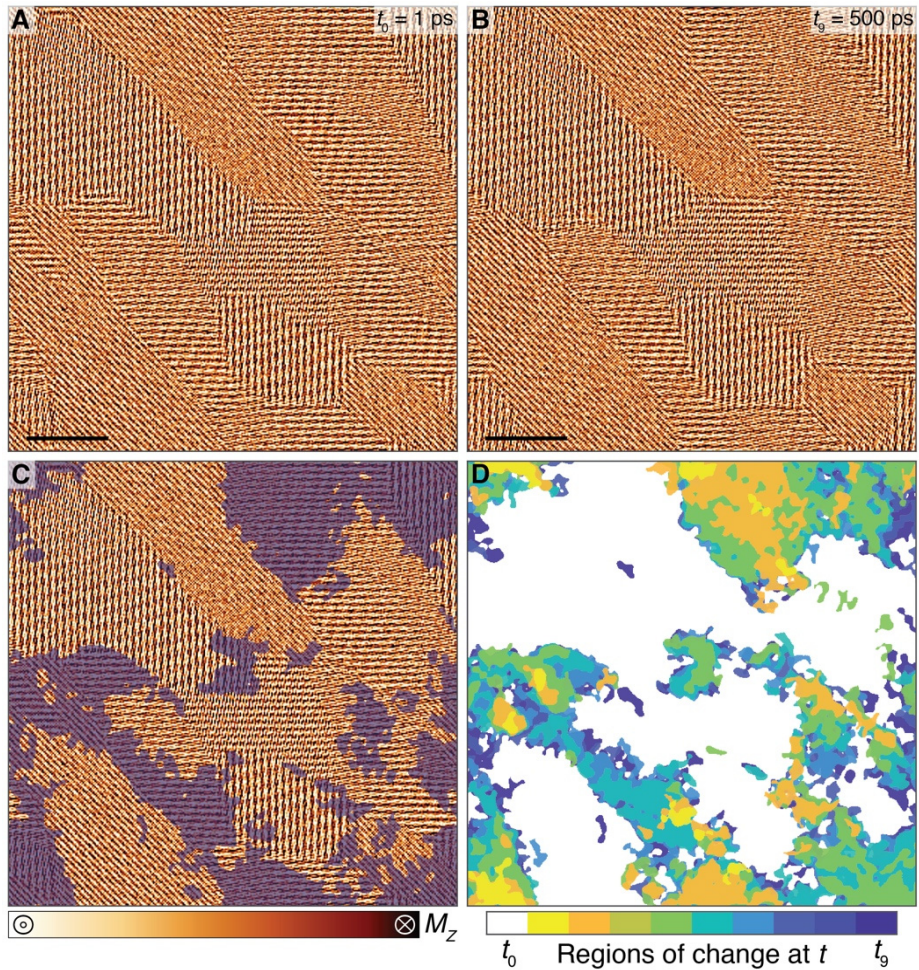

**Fig. 5. Observation of dynamic heterogeneity and nucleation in atomistic spin dynamics simulations. (A)** Simulated magnetization image at $t_0 = 1$ ps showing local magnetic patterns of multi-$Q$ character. **(B)** Magnetization image at a later time $t_7 = 500$ ps. **(C)** Map of regions that have evolved over the simulated time (blue) overlayed over the image at $t_0$ as reference revealing spatially heterogeneous dynamics. **(D)** Regions of change as a function of time (image at time $t_i$ compared to initial state at $t_0$, $\Delta t = 70$ ps) indicate a coexistence of stable (white) and changing (colored) regions that nucleate randomly and then grow over time (scalebar: 20 nm).

# Supplementary Materials for

## Dynamic heterogeneity in the self-induced spin glass state of elemental neodymium

L. Niggli, J. H. Strik, Z. Liu, A. Bergman, M. I. Katsnelson, D. Wegner, A. A. Khajetoorians*

*Corresponding author: a.khajetoorians@science.ru.nl

**The PDF file includes:**

    Materials and Methods
    Supplementary Text
    Figs. S1 to S12
    References (*43-46*)

**Other Supplementary Materials for this manuscript include the following:**

    Movies S1 to S2

**Materials and methods**
Nd(0001) islands were epitaxially grown on a cleaned W(110) substrate using the Stranski–Krastanov mode (see Ref. (*24*) for details on sample growth). The thicknesses of the four studied islands were between 70-180 ML and had a lateral size of at least 300 nm. As previously shown, these islands are thus bulk-like (*24*). The experimental studies were performed in a home-built, ultra-high vacuum system, which operates at a base temperature of 1.3 K with magnetic fields up to 9 T perpendicular to the sample. The SP-STM measurements were acquired with an antiferromagnetic Cr tip and prepared with out-of-plane spin-polarization. Magnetization images were produced by subtracting the majority ($V_S$ = -150 mV) and the minority ($V_S$ = 200 mV) SP-STM images (detailed description in Ref. (*24*)), and corresponding $Q$-space images were produced by computing the FFT. The image data processing was performed using MATLAB.

The similarity of two filtered $S(Q)$'s was quantified using the Jensen-Shannon (JS) divergence $\mathcal{D}_{JS}$, which is an established mathematical tool in information theory and possesses all the properties of a topological distance or metric (*43, 44*). The difference between two normalized discrete distributions, $P$ and $Q$ was calculated as follows:

$$\mathcal{D}_{JS}(P,Q) = \frac{1}{2}\sum_i \left(P_i \log\left(\frac{P_i}{M_i}\right) + Q_i \log\left(\frac{Q_i}{M_i}\right)\right) \text{ where } M_i = \frac{P+Q}{2} \quad \text{S1}$$

An additional property of $\mathcal{D}_{JS}$ is that its values are in between 0 and 1 when a base-2 logarithm is used and these values are reached when $P$ and $Q$ are identical ($\mathcal{D}_{JS}$ = 0) or two completely different distributions ($\mathcal{D}_{JS}$ = 1).

**Theory**
All simulations were performed using the atomistic spin dynamics software UppASD (*45*), with full algorithmic details provided in Ref. (*46*). The dhcp Nd system was modelled using the same spin Hamiltonian as described in Ref. (*24*), specifically a Heisenberg Hamiltonian parameterized with scalar Heisenberg exchange interactions derived from density functional theory (DFT). No anisotropic contributions, such as on-site magnetocrystalline anisotropy or anisotropic exchange interactions were included in the Hamiltonian. The magnetization snapshots presented in Fig. 5 were obtained by simulating a slab containing 300 × 300 × 4 dhcp unit cells and periodic boundary conditions were assumed in the x and y directions. To ensure that the flat geometry considered in this work did not alter the dynamics earlier observed for thicker samples (*24*), we confirmed by studies of the two-time autocorrelation function (see S10 for results and details) that these systems reproduce the glassy behavior as reported in Ref. (*24*).

**Supplementary Text**

**S1. Imaging the real space magnetization at *T* = 1.3 K**
Images as shown for example in Fig. 2A-C and Fig. 4A-B take 27 hours. In Fig. S1, we show repeated measurements over the course of 27 hours, of the same area at *T* = 1.3 K. We observed stable magnetic patterns with very few changes in subsequent images, as can also be confirmed in the respected $S(Q)$-maps (Figs. S1E-H, J, L). In this way, we froze out the aging dynamics at *T* = 1.3 K on the experimental time scale.

**S2. Filtering *S*(*Q*)-maps to extract spectral weight of *Q*-pockets and error bars**

In order to quantitatively compare $S(\boldsymbol{Q})$-maps in the self-induced spin glass state and investigate the spatial and radial distribution of $Q$-pockets, we developed a filtering method. We started by using the $S(\boldsymbol{Q})$-map measured in the long-range ordered state ($T_g < T < T_N$), in order to estimate the background noise. Fig. S2A shows such an image measured at $T = 9.3$ K that is characterized by a small number of sharp $Q$-spots on top of an average noise floor possessing a radial dependence. First, we filtered out the spectral weight of the $Q$-spots as shown in Fig. S2B and calculated a radial average $\mu_r(q)$ over the remaining noise in the $S(\boldsymbol{Q})$-map. We defined a background threshold $\tau_{\text{bg}}(q) = \mu_r(q) + \sigma_r(q)$, where $\sigma_r(q)$ denotes the radial standard deviation, that is displayed in Fig. S2C. Subtracting $\tau_{\text{bg}}(q)$ from a typical $S(\boldsymbol{Q})$-map in the self-induced spin glass phase (Fig. S2D) removes the radially dependent part of the noise as can be seen in Fig. S2E. Next, we calculated the histogram of the background subtracted $S(\boldsymbol{Q})$-map and fitted the distribution with a logistic fit $f(I, \mu, \sigma)$ due to the pronounced tails compared to a gaussian distribution:

$$f(I, \mu, \sigma) = \frac{1}{4\sigma} \text{sech}^2\left(\frac{I - \mu}{2\sigma}\right) \rightarrow \tau_1 = \mu + n\sigma \qquad \text{S2}$$

where $I$ denotes the intensity of the $S(\boldsymbol{Q})$-map. Based on the fit we defined another threshold $\tau_1$ to filter out the remaining noise ($\tau_1$ denotes a global value with no radial dependence on $q$). We adjusted $n = 1\text{-}3.5$ depending on the size and cleanness of the different samples, which affects the signal-to-noise ratio to optimize the filtering. The resulting binary $S(\boldsymbol{Q})$-map is shown in Fig. S2G. We lastly identified the connected regions of spectral weight using Matlab's imaging tools: *bwconncomp* to find connected components in the binary image, *regionprops* to assess their properties (for example area) and *imfill* to fill holes in the image. The result is shown in Fig. S2H and exhibits distinct pockets along the high symmetry axes together with some weight at $Q = 0$ and off-axes. The latter often results from higher order effects intrinsic to the Fourier transformation and hence we focused in our investigation solely on spectral weight along the high symmetry axes as shown in Fig. S2I. Such binarized $S(\boldsymbol{Q})$-maps represent the relevant spectral weight of the $Q$-pockets after filtering out the spectral weight related to noise. Hence, we define the error bars in Fig. 3C-D of the main text based on the JS-divergence of the filtered out noise.

**S3. Cooling rate**
We investigated the influence of the cooling rate on the reinitialization effect. To this end, we compared our typical cooling rate of ~0.8-1 K/min, which we defined as the standard rate in our measurements, to a slower cooling rate of ~0.05 K/min. Fig. S3A-B displays the real space magnetization images along with the corresponding $Q$-space images (Fig. S3C-D) obtained after reinitializing the sample using the two aforementioned cooling rates. There is no discernable difference, when comparing the radial distribution of the $Q$-pockets (Fig. S3E) as well as their spatial distribution in the $S(\boldsymbol{Q})$-maps.

**S4. $S(\boldsymbol{Q})$-evolution and reinitialization for different samples**
We applied the dynamical procedure outlined in the main text to four different samples. These samples had different levels of intrinsic defects (surface defect concentration in the range of ~ 0.003-0.006 ML ). Fig. S4 displays the magnetization images together with the corresponding filtered $S(\boldsymbol{Q})$-maps for one set of initial, final and reinitialized states for three other samples than the one in the main text (Fig. 2). We note that the relative alignments of the high symmetry axes for the different samples, varied due to different orientations of the scan frame used between different samples. All the investigated samples exhibit an evolution of $S(Q)$ from an initial towards a final state, where the $Q_A$-pocket moved radially outwards, while the $Q_B$- and $Q_C$-pockets shifted radially inwards as can be seen in the $Q$-distributions shown in Fig. S5G,N and U. In addition, when applying step (III) of the dynamical procedure, all measured samples exhibited a reinitialization of $S(Q)$ in terms of the $Q$-pocket position. We note that the specific $S(Q)$'s between the different samples exhibit variations in the extent and shape of the $Q$-pockets, while the evolution and reinitialization of $S(Q)$ are reproduced.

## S5. Identifying slow and fast dynamics in real space magnetization images

In order to resolve dynamic heterogeneity in real space, we compared local areas of measured large-scale magnetization images before ($t_i$) and after ($t_{i+1}$) inducing magnetization dynamics with a $B$-field cycle. The images were divided into a grid of boxes of side length $L$ after ensuring that both images display the same area using defects as a reference. The first criterion to identify local dynamics concerned the specific configuration of the local pattern: Each box was examined by eye to identify the dominant periodicities making up the local order (see examples in Fig. S5A). Subsequently, it was checked if the detected patterns at time $t_i$ and $t_{i+1}$ match. If the same periodicities were present before and after the perturbation, the box was labeled as possessing slow dynamics (Fig. S5A right), while a change of the periodicities illustrated fast dynamics (Fig. S5A left). The second criterion concerned the area a pattern occupies, as seen in Fig. S5B: if a box showed a partial change of pattern in much less (much more) than half of its area, the box was considered slow (fast). Changes amounting to approximately half of a box were labeled as undetermined. Fig. S5C-F display various examples of boxes to illustrate the application of the two criteria outlined above also in cases that are more ambiguous. Patterns judged as slow dynamics may exhibit changes on a small scale, like a $Q_A$ stripe doubling (Fig. S5C left) or a phase shift of the pattern as a whole (Fig. S5C right) that do not affect the present dominant $Q$-states. Besides a complete change of $Q$-states constituting a local pattern, the following cases were also judged as fast dynamics: (i) An additional $Q$-state emerges on top of a pattern (Fig. S5D), (ii) periodicity variation within a $Q$-pocket (Fig. S5E) and (iii) Straightening of a present $Q$-state (Fig. S5F) or combinations of any of these options.

## S6. Dynamic heterogeneity on sample with lower defect density

The sample used in Fig. 4 of the main text to demonstrate DH is different than the sample used in Figs. 1-3 of the main text. The latter possesses a lower defect density and we applied $B$-field cycles to higher magnetic fields. Here, we analyzed the spatial distribution of slow and fast dynamics during three magnetic field cycles to $B = 7$ T on the cleaner sample of Figs. 1-3. Fig. S8 displays the resultant magnetization images along with spatial maps of the differently identified dynamics. Figs. S8G-I show the same qualitative trend, namely a coexistence of slow and fast dynamics for all field cycles.

## S7. Dependence of slow/fast dynamics on sampled area

We explored the dependence of slow and fast dynamics on the box size $L$. Fig. S7A illustrates a local area imaged with a reducing box size, which allows to characterize the local pattern on different length scales. There are two limiting cases: (i) large box sizes, e.g. larger than the typical pattern size, increase the probability to find different dynamics (see largest box in Fig. S7A), (ii) small box sizes, i.e. on the order of the $Q_A$ pocket periodicities, reveal on the one hand small variations in the magnetization within a pattern (see smallest box in Fig. S7A), but prevent the identification of the magnetic pattern and its dominant $Q$-states. Fig. S7B-H shows the identification of slow and fast dynamics for diminishing box sizes $L$. For all chosen $L$, both slow and fast dynamics can be found, however the amount of boxes exhibiting undetermined dynamics (transparent boxes) strongly depends on the box size. The percentage of boxes possessing slow, fast or undetermined dynamics is shown in Fig. S7I and indicates the optimal box size for sorting predominantly into slow and fast categories lies around $L = 12$ nm.

**S8. Decomposition of simulated real space magnetization into magnetic sublattices**
The double hexagonal closed packed crystal structure of Nd(0001) with an ABACA stacking results in two distinct sublattices, which possess different local symmetries: the A layer is sandwiched between two different layers and hence exhibits local cubic symmetry, while the B and C layers are surrounded by the same layer and thus exhibit locally a hexagonal symmetry. They are accordingly referred to as cubic (cub) and hexagonal (hex) sublattices. The effect of these different environments on the exchange interaction was discussed in Refs. (*24, 25*). While the simulated magnetization images in the main text (Fig. 5) display the combined signal of both sublattices, Fig. S8 shows the decomposition of the magnetization into the hex and cub sublattices for $t_0$ (A-B) and $t_7$ (C-D) respectively. Both sublattices show short-range order yet with different overall periodicities. Following the local orders for the hex and cub sublattices between $t_0$ and $t_7$ reveals spatially heterogeneous dynamics in the form of stable and changing regions similar to the behavior of the total simulated magnetization images shown in Fig. 5 of the main text.

**S9. Representation of simulated magnetization textures**
Figure 5 in the main text as well as Figs. S8 and S12 display simulated magnetic configurations for dhcp Nd represented as false-color images. The configurations represented in these figures consist of $300 \times 300 \times 4$ unit cells of Nd, with a hexagonal simulation box. To simplify the analysis, we have cropped the area and plotted the magnetic textures in a square representation. This representation corresponds to a coordinate system expressed in the lattice vectors of the actual system and has the benefit that the magnetic texture can readily be expressed on a uniform grid or as elements in a square matrix. As a result of this transformation, all simulated figures display a skewed realization of the real-space spin texture. This linear transformation of the spin textures does not affect the analysis of local dynamics performed in this work but should be noted in case of comparisons with earlier works on the system as done in Refs. (*24, 25*). An example on how the square representation transfers to the un-skewed real-space representation can be seen in Fig. S9.

**S10. Subtraction method to extract region of changes**
We applied the following method to quantify regions in the simulated magnetization images that exhibited a change of the local order between a time $t$ and the initial time $t_0$: Fig. S10A-B displays two example magnetization images ((A) initial state at time $t_0$, (B) evolved state at time $t_3$) that were to be compared in terms of their spin configurations. First, we subtracted the two images (see Fig. S10C), which revealed regions in space where the magnetic contrast approximately vanished. Next, we filtered out those spatial regions possessing a magnetization value close to zero by choosing a threshold that extracts the top 30 % of the extremal magnetization values. The resulting binary image is shown in Fig. S10D and clearly shows regions of accumulated signal as well as regions possessing a lower density of extremal magnetization values. We applied a gaussian filter and another filtering threshold ($\tau_c$ = 0.25) to extract the regions of maximal difference in (D) as can be seen in Fig. S10E. In the last step, we filled holes within connected regions using MATLAB's *imfill* function and also neglected the smallest connected regions. Thereby, we obtained a binary image quantifies the regions of change between two time steps (Fig. S10F). We note that this simplistic analysis does not give information about the nature of the change, i.e. it does not distinguish between a phase shift or a complete change of the local order.

**S11. Statistics of simulated local dynamics.**
In the main text, the magnetization dynamics from a single ASD simulation are used to demonstrated DH and the spatially resolved dynamics in dhcp Nd. To demonstrate that this one simulation is not a statistical outlier or exception, we performed a series of simulations on dhcp Nd for the same system size (300 × 300 × 4 unit cells) where starting states as well as the sequence of random numbers that populate the heat-bath in our Langevin-based simulations (for more details on stochastic ASD simulations see Ref. (*46*)), Regardless of initial state or random-number streams, we observed similar local dynamics for all simulated samples, as can be seen in Fig. S11. Different choices of the Gilbert damping parameter were also considered with the same, positive, outcome for all cases (data not shown).

**S12. Autocorrelation function**
In Ref. (*24*), the simulated autocorrelation function
$$C(t, t_w) = \langle \boldsymbol{m}_i(t_w)\boldsymbol{m}_i(t + t_w)\rangle \qquad \text{S3}$$
showed glassy behavior for bulk dhcp Nd. For illustrative purposes we have used a different simulation cell for the ASD simulations in this work (here we use 300 × 300 × 4 unit cells compared to 32 × 32 × 32 unit cells used in Ref. (*24*)). To ensure that we capture glassy dynamics for the systems simulated here, we calculated the autocorrelation $C(t, t_w)$. For comparison we also simulated an Edwards-Anderson model (EA) for the same 300 × 300 × 4 geometry. These autocorrelation functions are shown in Fig. S12 and can be compared to previous results in Ref. (*24*). We can conclude that for the systems sizes considered in this study we obtain a similar, but not identical, relaxation behavior for both the dhcp Nd and the EA model systems.

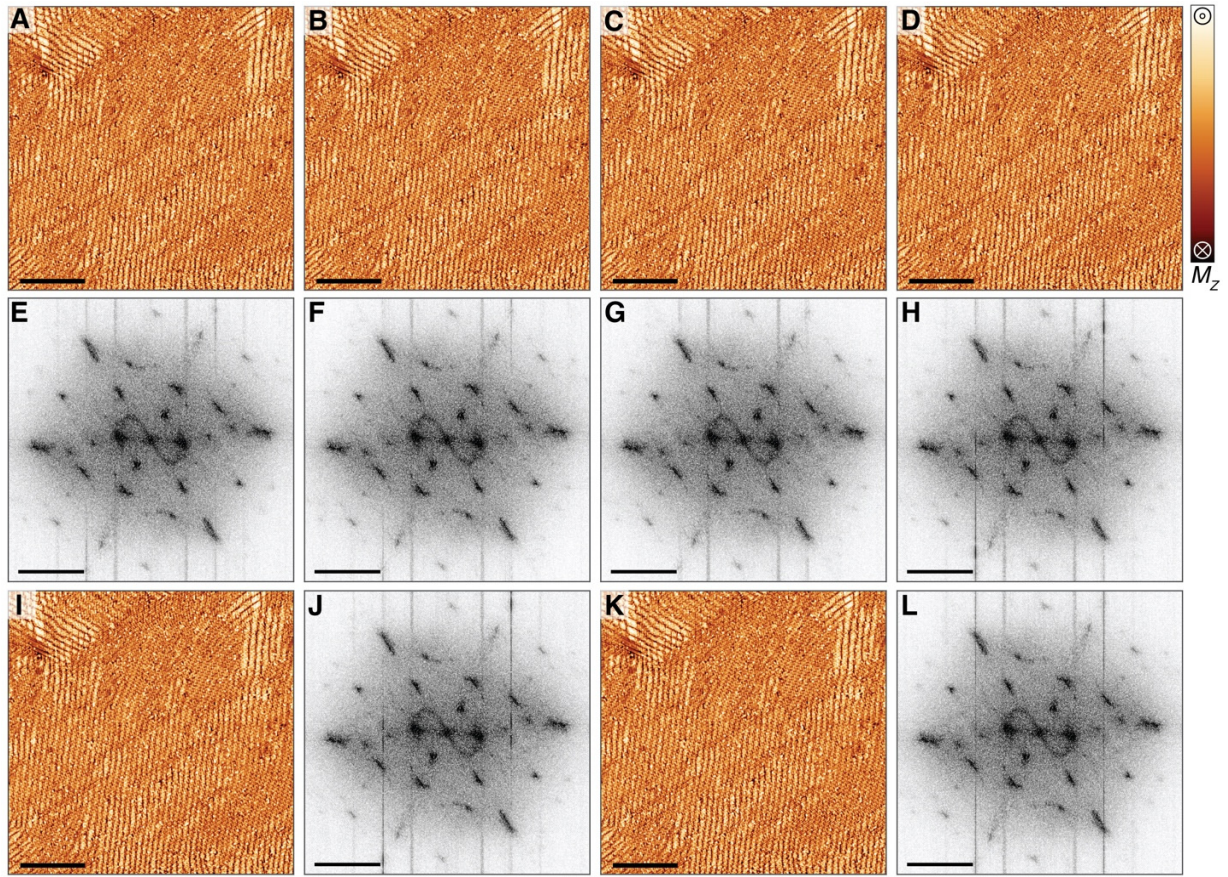

**Fig. S1. Frozen real space magnetization at $T$ = 1.3 K. (A-D, I, K)** Large scale magnetization images of roughly the same area as in Fig. 2 of the main text measured consecutively at $T$ = 1.3 K and $B$ = 0 T (scan direction down, up, down, etc.), where the duration of one image corresponds to ~ 4.5 h, show no change of magnetic patterns over ~ 27 h. **(E-H, J, L)** $S(Q)$-maps corresponding to the images in the row above (E-H) or the images to the left **(J, L)** (image parameters: $I_t$ = 200 pA; scale bar for all $M(r)$ images: 50 nm, colorbar for all $M(r)$ in the top right, color scale for $S(Q)$-images is a linear gradient).

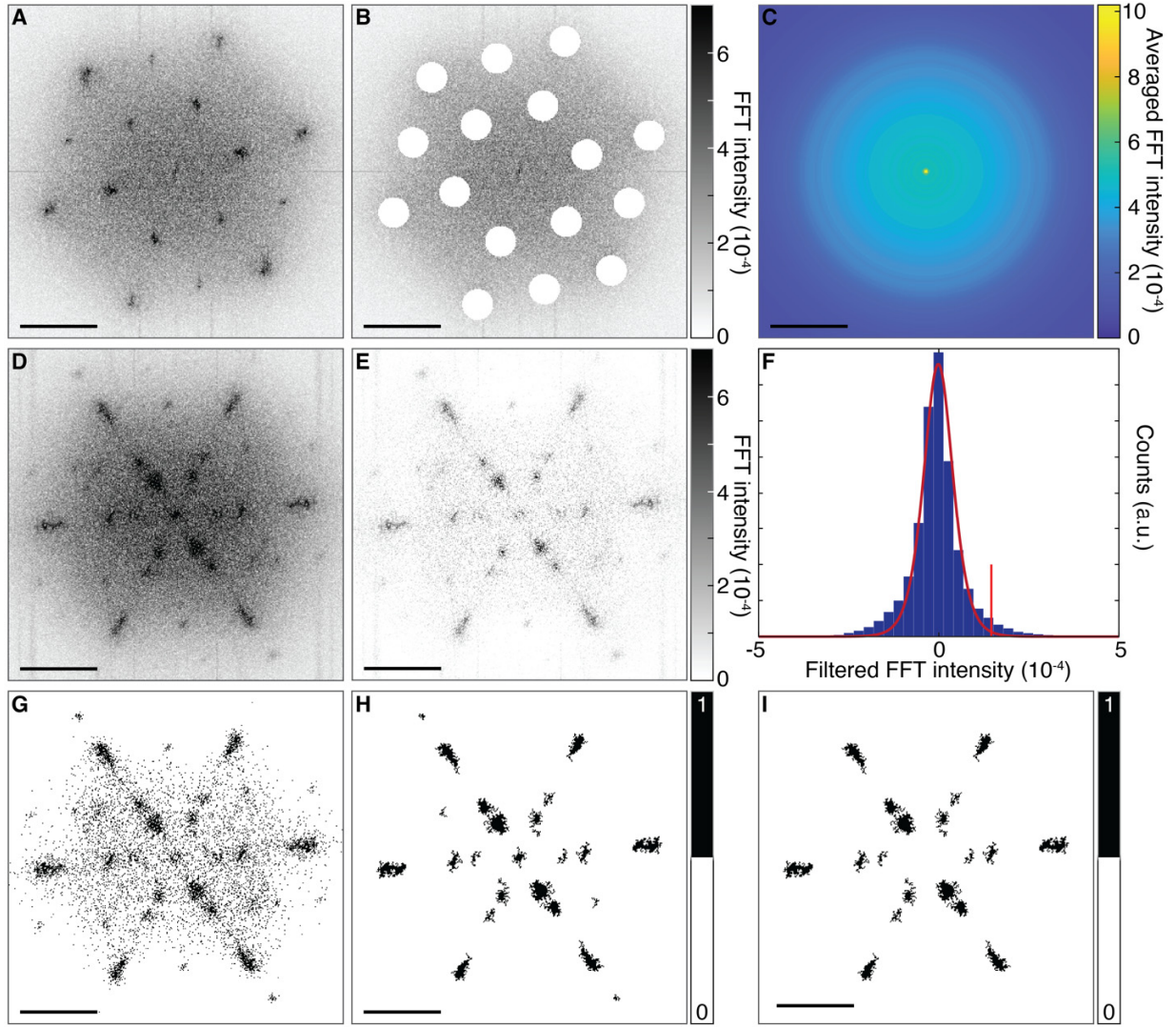

**Fig. S2. Systematic method to extract the $Q$-pockets from $S(Q)$-maps. (A)** representative $S(Q)$-map in the multi-$Q$ phase at $T = 9.3$ K distinguished by sharp $Q$-spots. **(B)** $S(Q)$-map of **(A)** containing solely noise after removing spectral weight of $Q$-spots. **(C)** Radially dependent background threshold $\tau_{bg}$ extracted by radial averaging of intensity of **(C)**. **(D)** Representative $S(Q)$-map in the self-induced spin glass state at $T = 1.3$ K characterized by broadly distributed $Q$-pockets. **(E)** Resulting $S(Q)$-map after subtracting $\tau_{bg}$ from image shown in **(D)**. **(F)** Histogram of the FFT intensity of (E) as well as a logistic fit to the distribution (red line) that yields threshold $\tau_1$. **(G)** Resulting binary image after applying the threshold $\tau_1$ to **(E)**. **(H)** Binary image containing only connected regions of a significant size **(I)** binary image exhibiting the spectral weight of the $Q$-pockets along the high symmetry axes after removing weight at $Q = 0$ and off-axis (scale bar for all images: 3 nm$^{-1}$).

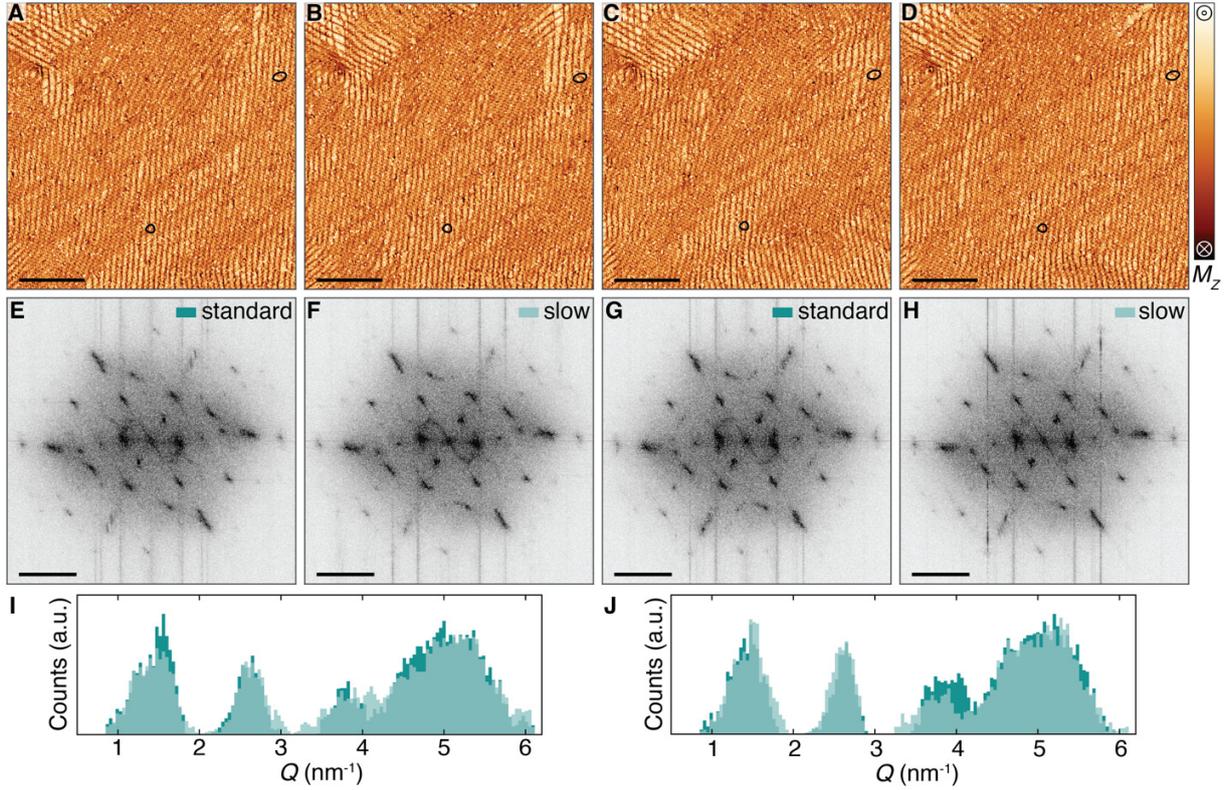

**Fig. S3. Cooling rate dependence. (A-B)** Magnetization images after reinitialization using our standard rate of ~ 1 K/min **(A)** and a slower cooling rate of ~ 0.05 K/min **(B)** to cool from the multi-$Q$ phase into self-induced spin glass phase down to $T$ = 1.3 K (black circles mark two defects as a common reference for all images). **(C-D)** Repeated measurement using our standard **(C)** and slow **(D)** cooling rate. **(E-H)** $S(Q)$-maps corresponding to the magnetization images in the row above. **(I-J)** Radial $Q$-distributions of the $S(Q)$-maps **(E-F)** and **(G-H)** respectively remain unaffected by the cooling rate (image parameters: $I_t$ = 200 pA; scale bar $M(r)$: 50 nm, scale bar $S(Q)$-maps: 3 nm$^{-1}$, color scale for $S(Q)$-images is a linear gradient).

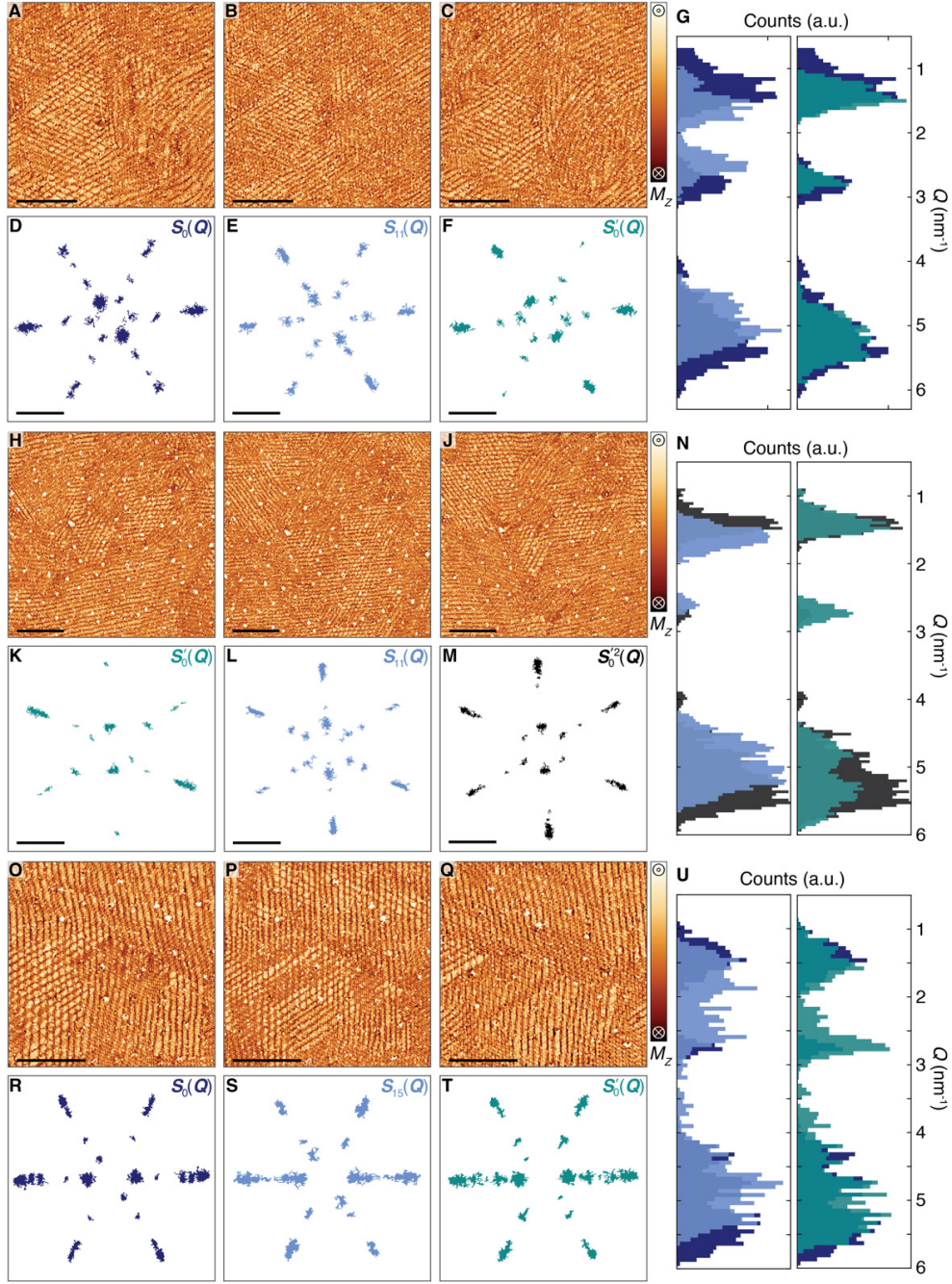

**Fig. S4. $S(Q)$-evolution and reinitialization on different samples. (A-C, H-J, O-Q)** Magnetization images at $T$ = 1.3 K for three different samples: sample #1 **(A-C)**, sample #2 **(H-J)** and sample #3 **(O-Q)**, where left image displays initial state, middle image shows a final state and right image exhibits a reinitialized state for each sample (sample #2 compares two reinitialized states). **(D-F, K-M, R-T)** Magnetic structure factors $S(Q)$ corresponding to each magnetization image shown in the row above. **(G, N, U)** Radial $Q$-distributions of the filtered $S(Q)$'s displaying (left) a comparison of initial and final states and (right) a comparison of initial and reinitialized states (image parameters: $I_t$ = 200 pA; scale bar $M(r)$: 50 nm, scale bar $S(Q)$-maps: 3 nm$^{-1}$).

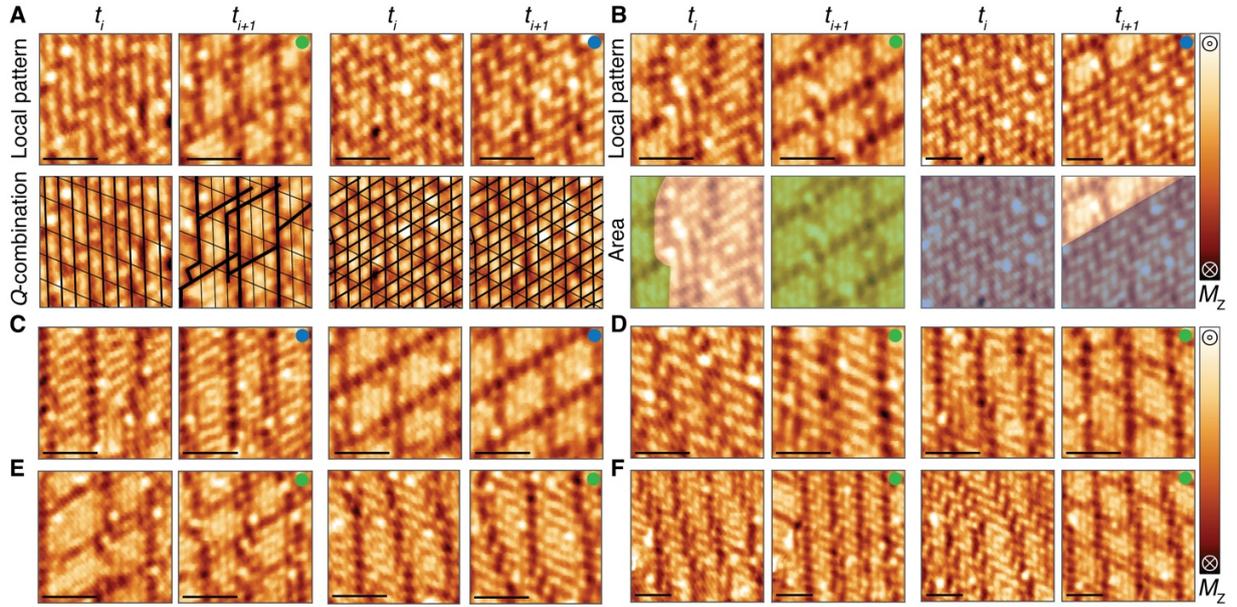

**Fig. S5. Criteria for identification of slow/fast dynamics.** (**A**) (top row) sets of real space images of two magnetic patterns before ($t = t_i$) and after ($t = t_{i+1}$) a *B*-field cycle to induce dynamics, (bottom row) decomposition of pattern into *Q*-states by eye. (**B**) (Top row) sets of real space magnetization images before ($t = t_i$) and after ($t = t_{i+1}$) a *B*-field cycle to induce dynamics, (bottom row) area of distinct patterns identified by eye (**C**). (**D**) Two examples of patterns that exhibit an additional *Q*-state after the applied *B*-field cycle. (**E**) Two examples of patterns that show a variation of the periodicity within a *Q*-pocket. (**F**) Two examples of patterns that exhibit a straightening of the $Q_A$-pocket periodicity after field cycling (image parameters: $I_t$ = 200 pA; scale bar for all images: 5 nm; colored dot in right corner labels slow (blue) and fast (green) dynamics).

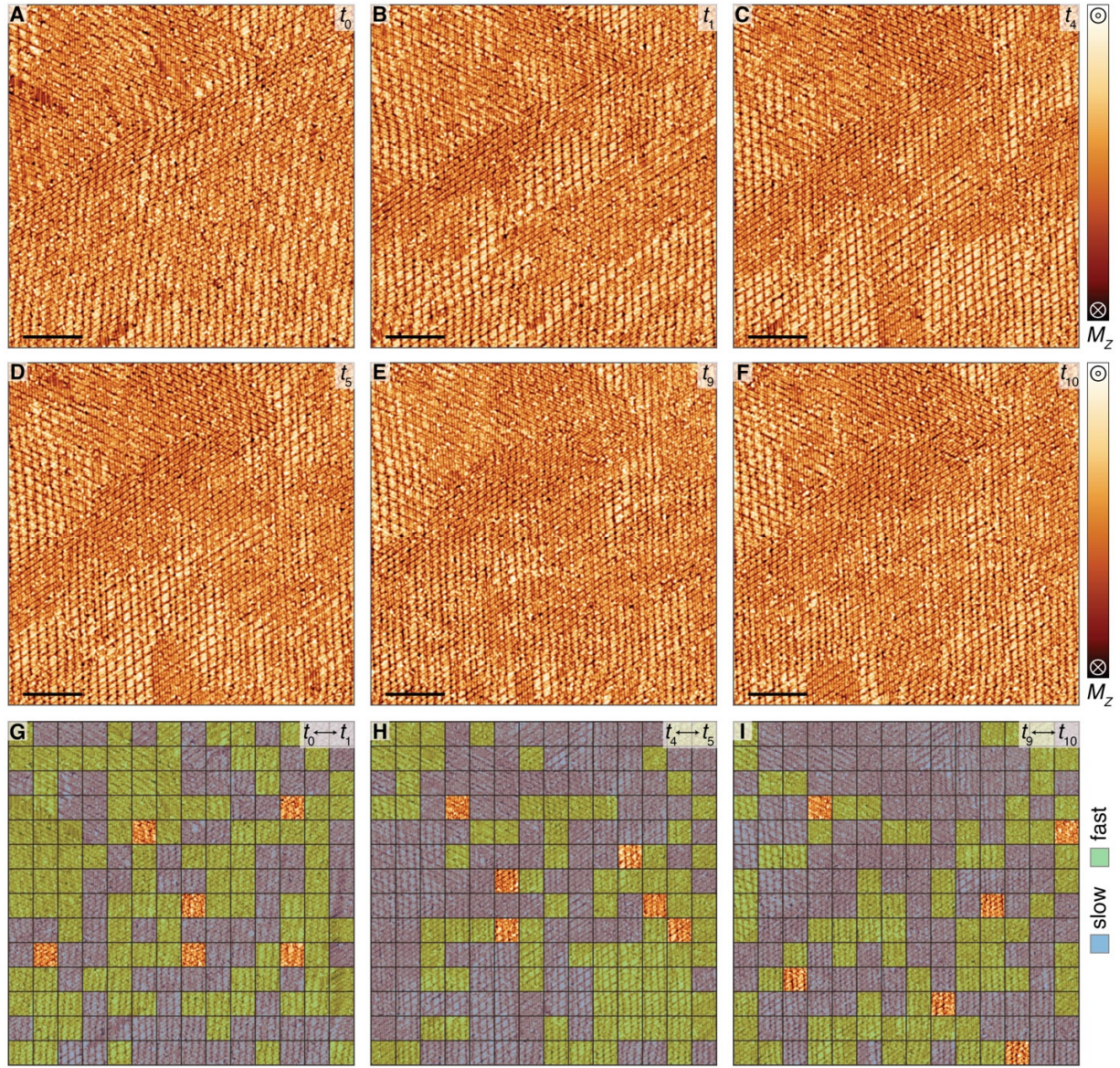

**Fig. S6. DH on a sample with a lower defect density. (A-F)** Magnetization images for three applied magnetic field cycles on the sample used in Figs. 1-3 of the main text: **(A)** initial state after ZFC, **(B)** first field cycle with $\Delta B = 7$ T. **(C-D)** Images before and after fifth field cycle with $\Delta B = 7$ T. **(E-F)** Images before and after ninth field cycle with $\Delta B = 7$ T. **(G-I)** Qualitative spatial map of the distribution of slow (blue) and fast (green) dynamics between **(G)** $t_0$ and $t_1$, **(H)** $t_4$ and $t_5$ and **(I)** $t_9$ and $t_{10}$ illustrates the presence of DH for all applied field cycles to a higher magnetic field on a sample with a lower defect concentration (image parameters: $I_t = 200$ pA; scale bar $M(r)$: 30 nm).

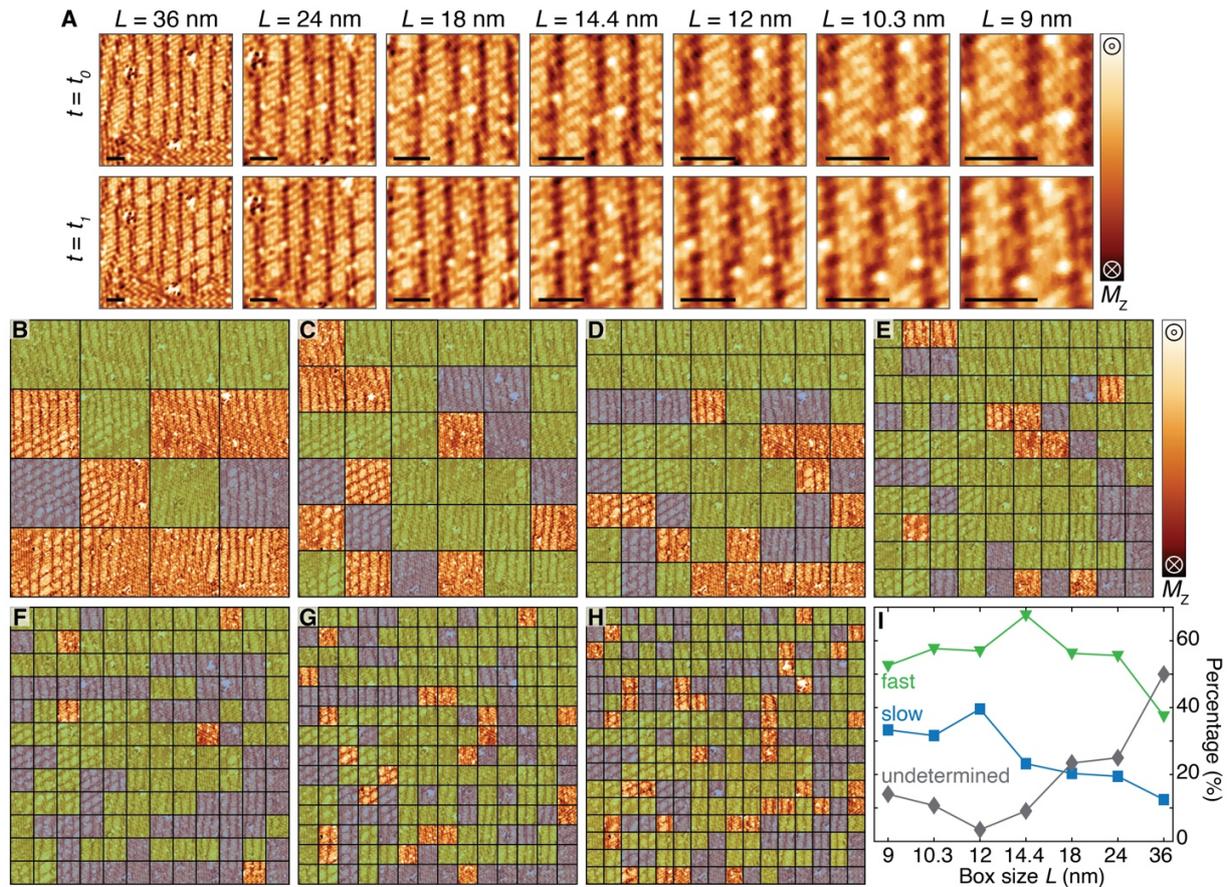

**Fig. S7. Box size dependence for identification of spatially heterogeneous dynamics. (A)** specific area of a magnetization image (Fig. 4A in the main text) displayed with different box sizes $L$ before and after a B-field cycle to 2 T. **(B-H)** Spatial map of identified slow (blue) and fast (green) dynamics for increasing box sizes $L$. **(I)** Percentage of total boxes in **(B-H)** exhibiting slow (blue), fast (green) and undetermined (grey) dynamics (image parameters: $I_t$ = 200 pA; scale bar for **(A)**: 5 nm, image size **(B-H)**: 144 nm).

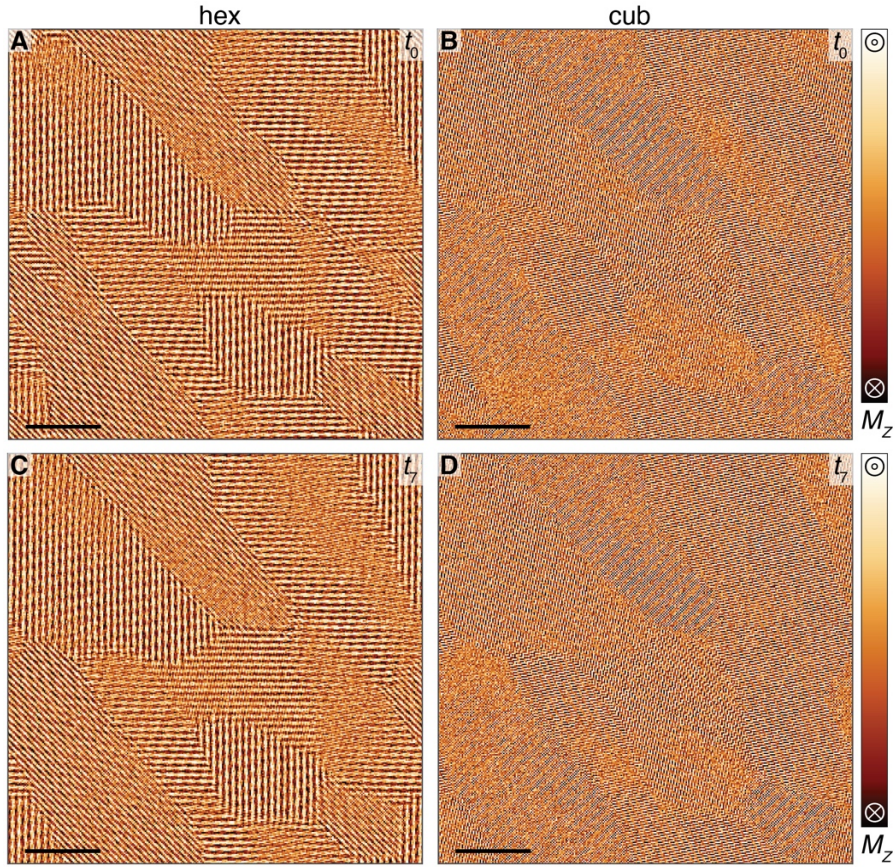

**Fig. S8. Magnetization decomposed into the distinct sublattices. (A-B)** Magnetization images of the top **(A)** hexagonal layer and **(B)** cubic layer at the beginning of the simulation (time $t_0$) illustrate the varying periodicities present. **(C-D)** Magnetization images of the top **(C)** hexagonal layer and **(D)** cubic layer at the end of the simulated time $t_7$ revealing a coexistence of stable and changing regions in both sublattices (scale bar: 20 nm).

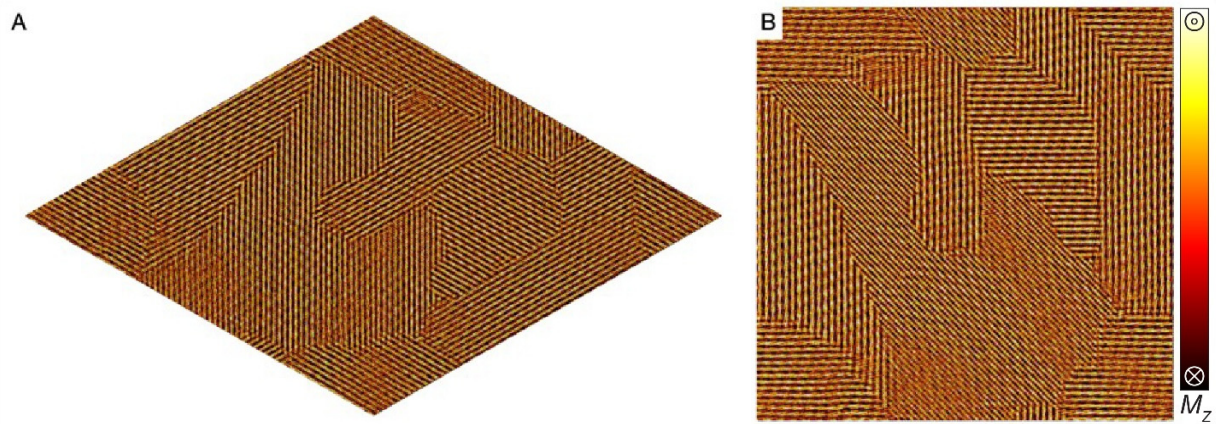

**Fig. S9. Raw and skewed simulation data. (A)** undistorted texture of one atomic layer (hexagonal) in a 300 × 300 × 4 unit cell simulation box of dhcp Nd. **(B)** magnetic texture of the same layer but projected onto a square image. Both images are represented as false-color images where the color indicates the out-of-plane component of the magnetization.

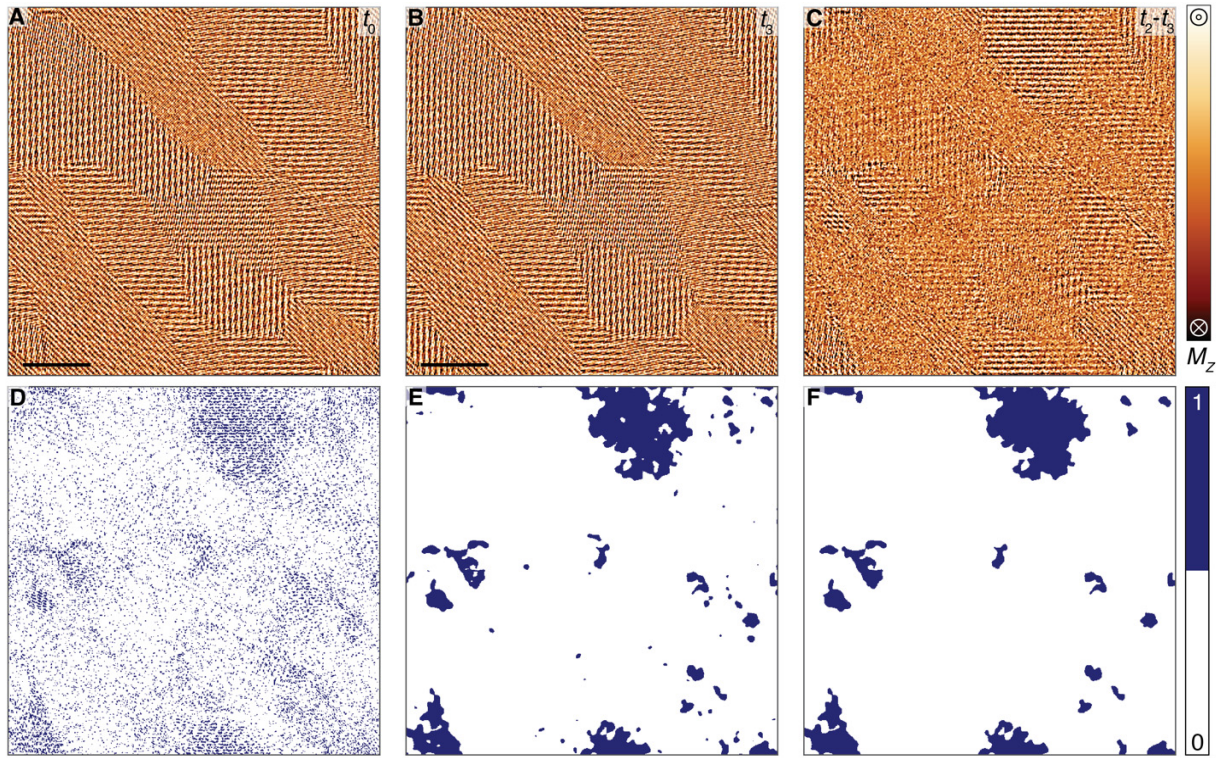

**Fig. S10. Filtering method to quantify regions of change in real space. (A-B)** Magnetization images at times $t_0$ and $t_3$. **(C)** Difference image of images shown in **(A)** and **(B)** exhibiting regions of strong as well as vanishing magnetic contrast. **(D)** Binary image representing positions of a different (blue) and stable (white) magnetization between $t_0$ and $t_3$ after extracting the positions of extremal contrast in **(C)**. **(E)** Identified connected regions of change in **(D)** based on a gaussian filter and additional threshold. **(F)** Finalized binary image representing stable (white) and changed (connected) regions (blue) between $t_0$ and $t_3$ after neglecting the smallest areas of change (scale bar: 20 nm).

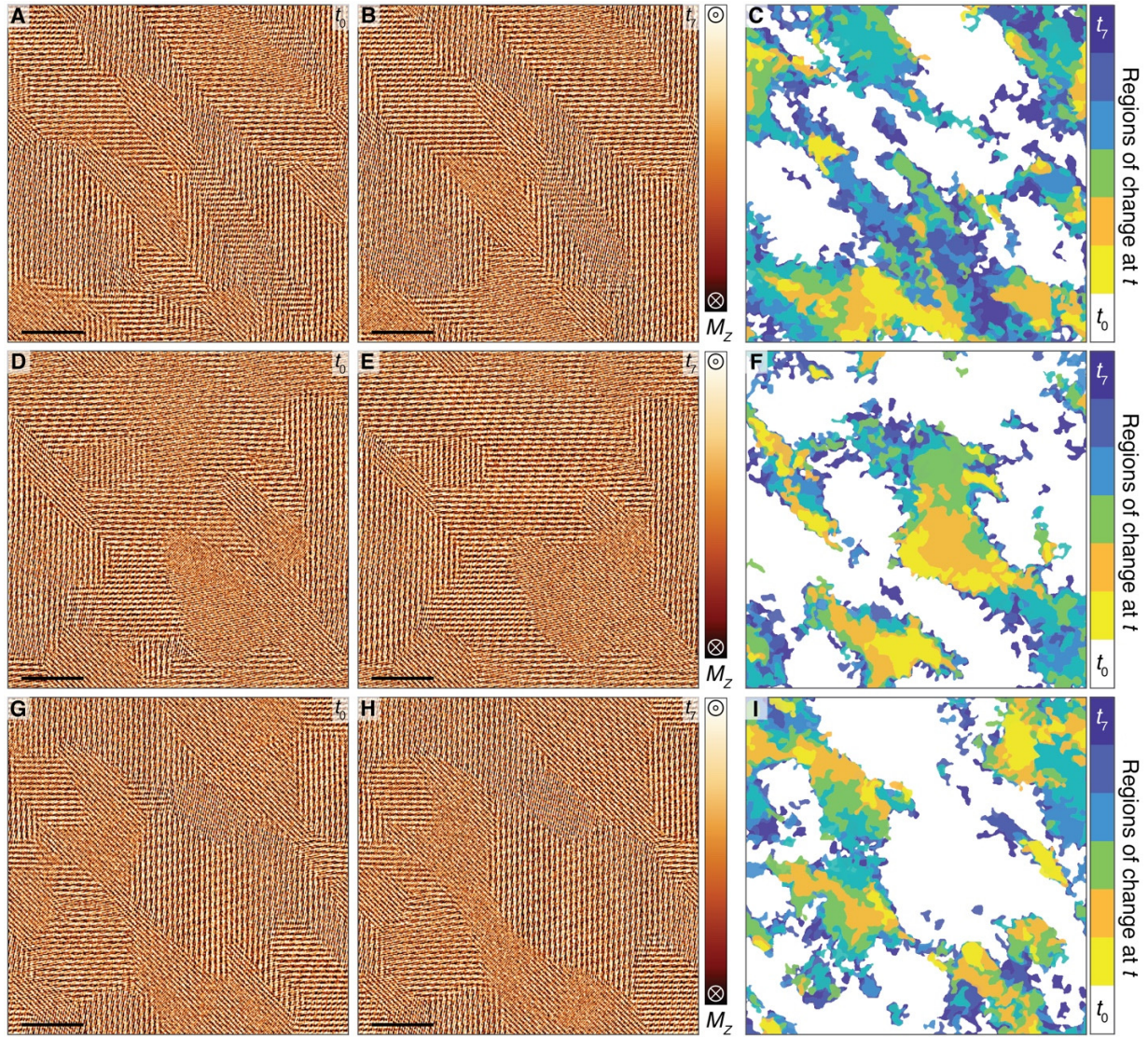

**Fig. S11. DH for distinct ASD simulations. (A-B)** Images showing snapshots of the magnetization at $t_0 = 1$ ps **(A)** and $t_7 = 500$ ps **(B)** for simulations of $300 \times 300 \times 4$ unit cells of dhcp Nd. **(C)** Corresponding regions of change as a function of time (image at time $t_i$ compared to initial state at $t_0$, $\Delta t = 70$ ps) indicate a coexistence of stable (white) and changing (colored) regions that nucleate at specific locations and then grow over time. **(D-F, G-I)** same analysis for simulation runs with different starting states and heat-baths showing similar nucleation behavior as well as a coexistence of stable and changing regions reminiscent of DH independent of the sample characteristics (scalebar: 20 nm).

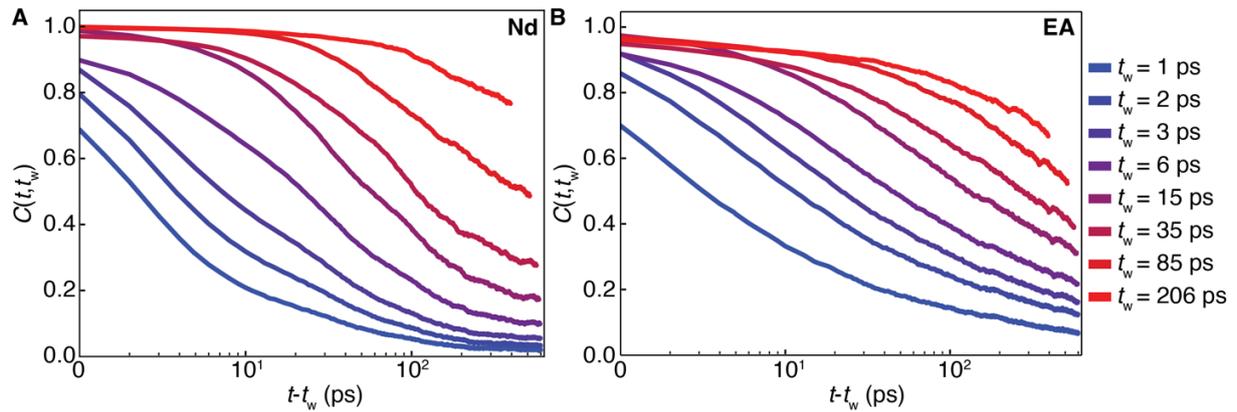

**Fig. S12. Autocorrelation function. (A)** Simulated autocorrelation function $C(t, t_w)$ (see text S12 for description of the methodology) for a $300 \times 300 \times 4$ unit cell simulation box of dhcp Nd. A glassy relaxation behavior is seen even though the waiting times $t_w$ span three orders of magnitude in time. **(B)** Simulated autocorrelation function $C(t, t_w)$ for a $300 \times 300 \times 4$ unit cell simulation for an Edwards-Anderson (EA) model on the dhcp lattice.

**Movie S1. Snapshots of the aging dynamics in real space.**

Spatially resolved magnetization images of the same area for 10 magnetic field cycles to $B = 7$ T illustrating the ongoing aging dynamics in elemental neodymium (image parameters: $I_t = 200$ pA).

**Movie S2. Simulated magnetization dynamics.**

Spatially resolved simulated magnetization images show aging dynamics in neodymium over a simulation time of $t = 500$ ps (time step between images $\Delta t = 20$ ps).